\documentclass[fleqn,usenatbib]{mnras}

\usepackage{newtxtext,newtxmath}

\usepackage[T1]{fontenc}

\DeclareRobustCommand{\VAN}[3]{#2}
\let\VANthebibliography\thebibliography
\def\thebibliography{\DeclareRobustCommand{\VAN}[3]{##3}\VANthebibliography}

\usepackage{graphicx}	
\usepackage{amsmath}	
\usepackage{subfig}     

\newsavebox{\measurebox}

\title[Blue-peak dominated lensed double-peaked LAE]{A double-peaked Lyman-$\mathbf{\alpha}$ emitter with a stronger blue peak multiply imaged by the galaxy cluster RXC~J0018.5+1626}

\author[Furtak et al.]{
Lukas J. Furtak,$^{1}$\thanks{E-mail: furtak@post.bgu.ac.il}
Ad\`{e}le Plat,$^{2}$
Adi Zitrin,$^{1}$
Micheal W. Topping,$^{2}$
Daniel P. Stark,$^{2}$
Victoria Strait,$^{3,\,4}$
\newauthor{St\'{e}phane Charlot,$^5$
Dan Coe,$^6$
Felipe Andrade-Santos,$^{7}$
Maru\v{s}a Brada\v{c},$^{8,\,9}$
Larry Bradley,$^6$}
\newauthor{Brian C. Lemaux,$^{9,\,10}$ and
Keren Sharon$^{11}$}
\\
$^{1}$Physics Department, Ben-Gurion University of the Negev, P.O. Box 653, Be’er-Sheva 84105, Israel\\
$^{2}$Steward Observatory, University of Arizona, 933 N Cherry Ave, Tucson, AZ 85721 USA\\
$^{3}$Cosmic Dawn Center (DAWN), Denmark\\
$^{4}$Niels Bohr Institute, University of Copenhagen, Jagtvej 128, DK-2200 Copenhagen N, Denmark\\
$^5$Sorbonne Universit\'{e}, CNRS UMR 7095, Institut d'Astrophysique de Paris, 98bis bvd Arago, 75014, Paris, France\\
$^6$Space Telescope Science Institute, 3700 San Martin Drive, Baltimore, MD 21218, USA\\
$^{7}$Center for Astrophysics \text{\textbar} Harvard \& Smithsonian, 60 Garden Street, Cambridge, MA 02138, USA\\
$^8$University of Ljubljana, Department of Mathematics and Physics, Jadranska ulica 19, SI-1000 Ljubljana, Slovenia\\
$^9$Department of Physics and Astronomy, University of California, Davis, One Shields Ave., Davis, CA 95616, USA\\
$^{10}$Gemini Observatory, NSF’s NOIRLab, 670 N. A’ohoku Place, Hilo, Hawai’i, 96720, USA\\
$^{11}$Department of Astronomy, University of Michigan, 1085 South University Drive, Ann Arbor, MI 48109, USA
}

\date{Accepted 2022 July 28; Revised 2022 July 27; Received 2022 April 21}

\pubyear{2022}

\begin{document}
\label{firstpage}
\pagerange{\pageref{firstpage}--\pageref{lastpage}}
\maketitle

\begin{abstract}
We report the discovery of a double-peaked Lyman-$\alpha$ (Ly$\alpha$) emitter (LAE) at $z=3.2177\pm0.0001$ in VLT/MUSE data. The galaxy is strongly lensed by the galaxy cluster RXC~J0018.5+1626 recently observed in the RELICS survey, and the double-peaked Ly$\alpha$ emission is clearly detected in the two counter images in the MUSE field-of-view. We measure a relatively high Ly$\alpha$ rest-frame equivalent width (EW) of $\mathrm{EW}_{\mathrm{Ly}\alpha,0}=(63\pm2)$\,\AA. Additional spectroscopy with \textit{Gemini}/GNIRS in the near-infrared (NIR) allows us to measure the H$\beta$, [\ion{O}{iii}]$\lambda$4959\AA\ and [\ion{O}{iii}]$\lambda$5007\AA\ emission lines, which show moderate rest-frame EWs of the order of a few $\sim10-100$\,\AA, an [\ion{O}{iii}]$\lambda$5007\AA/H$\beta$ ratio of $4.8\pm0.7$, and a lower limit on the [\ion{O}{iii}]/[\ion{O}{ii}] ratio of $>9.3$. The galaxy has very blue UV-continuum slopes of $\beta_{\mathrm{FUV}}=-2.23\pm0.06$ and $\beta_{\mathrm{NUV}}=-3.0\pm0.2$, and is magnified by factors $\mu\sim7-10$ in each of the two images, thus enabling a view into a low-mass ($M_{\star}\simeq10^{7.5}\,\mathrm{M}_{\odot}$) high-redshift galaxy analog. Notably, the blue peak of the Ly$\alpha$ profile is significantly stronger than the red peak, which suggests an inflow of matter and possibly very low \ion{H}{i} column densities in its circumgalactic gas. To the best of our knowledge, this is the first detection of such a Ly$\alpha$ profile. Combined with the high lensing magnification and image multiplicity, these properties make this galaxy a prime candidate for follow-up observations to search for LyC emission and constrain the LyC photon escape fraction.
\end{abstract}




\section{Introduction} \label{sec:intro}
The epoch of cosmic reionization (EoR), which took place at $z\sim5-15$ \citep[e.g.][]{stark10,becker15,planck16,planck20,banados19,zhu21,robertson21,bosman22}, saw one of the last major phase changes of the Universe as almost the entire neutral hydrogen content in the intergalactic medium (IGM) was reionized. The current consensus strongly suggests early star-forming galaxies, and in particular the faint low-mass population, as the sources that drove cosmic reionization \citep[e.g.][]{bunker10,bouwens11,bouwens15,robertson13,robertson15,mclure13,finkelstein15,atek15b}. The physics that govern the emission of ionizing photons $>13.6$\,eV by star-forming galaxies and, more importantly, their \textit{escape} into the surrounding IGM, however remain largely uncertain. Reionizing the entire IGM would require early star-forming galaxies to have escape fractions of ionizing radiation of $f_{\mathrm{esc,LyC}}>0.2$, on average \citep[e.g.][]{robertson13,naidu20}. Because of the opacity of the neutral IGM to Lyman-continuum (LyC; $\lambda_0<912$\,\AA) photons \citep[e.g.][]{madau95,inoue14}, it is essentially impossible to directly observe LyC emission of galaxies beyond $z\gtrsim4.5$ \citep{vanzella18}. We therefore need to indirectly infer the LyC escape fractions (and other emission properties) of high-redshift galaxies through signatures that \textit{can} be observed out to high redshifts. This is done by studying LyC leakage in low-redshift analogs of high-redshift galaxies, in particular in Lyman-$\alpha$ (Ly$\alpha$) emitters (LAEs).

The Ly$\alpha$ line originates from a resonant transition in the hydrogen atom. A connection between Ly$\alpha$ and LyC escape can therefore be expected theoretically. Indeed, Ly$\alpha$ photons scatter less in lower \ion{H}{i} column density regions which would also allow for LyC photons to escape \citep[e.g.][]{dijkstra16} and leave an imprint on the Ly$\alpha$ emission line profile \citep[][]{verhamme15,kakiichi21}. Also, a correlation between high escape fractions and high Ly$\alpha$ equivalent widths (EWs) is indicated by both simulations \citep[e.g.][]{maji22} and observations \citep[e.g.][]{verhamme17,marchi18,steidel18,izotov21,flury22b}. Numerous Ly$\alpha$ emitters (LAEs) have been observed at $z>6$ \citep[e.g.][]{pentericci14,pentericci18,schenker14,hu19,fuller20,goto21,wold22,endsley22}, some of them even with double-peaked Ly$\alpha$ profiles \citep[][]{hu16,matthee18,songaila18,meyer21,endsley22}. Note that the latter is perhaps surprising because the blue peak would have a higher chance of being scattered away by the increasingly neutral IGM at high redshifts. However, while the vast majority of confirmed low-redshift LyC leakers observed to date are strong LAEs, the opposite is not necessarily true: even in large samples of strong LAEs at moderate redshifts $z\sim3$, direct measurements of LyC escape fractions in systematic searches with ultra-deep \textit{Hubble Space Telescope} (HST) imaging and ground-based spectroscopy in ultra-violet (UV) wavelengths remain scarce \citep{bian20} and spurious in many cases \citep[e.g.][]{fletcher19,pahl21}. We therefore need to establish more distinct signatures of LyC leakage in LAEs in order to be able to more robustly identify and target LyC emitter candidates at both high and low redshifts \citep[e.g.][]{zackrisson11,zackrisson13,zackrisson17,reddy16,izotov18,izotov21,schaerer18,schaerer22,tang19,plat19,pahl20,matthee21,naidu22,flury22a,flury22b}.

In this work we present the serendipitous discovery of a unique case of double-peaked Ly$\alpha$ emission from a galaxy at $z\simeq3.218$. The galaxy is lensed by the galaxy cluster RXC~J0018.5+1626 (also known as CL0016+1609 or MACS~J0018.5+1626; e.g. \citealt{ebeling07}) at $z_{\mathrm{d}}\simeq0.546$, which was recently imaged with HST as a part of the \textit{Reionization Lensing Cluster Survey} (RELICS), building on previous HST observations \citep[][]{coe19}. We therefore dub the galaxy, RXC0018-LAE1. While double-peaked Ly$\alpha$ profiles are commonly observed in LAEs \citep[][]{leclercq17,izotov18,leclercq20,matthee21,kerutt22,naidu22}, RXC0018-LAE1 shows a peculiar Ly$\alpha$ profile with a significantly stronger blue peak, which could indicate low neutral gas covering fractions that could potentially also allow LyC photons to escape. In general, narrow Ly$\alpha$ peak separation \citep[e.g.][]{izotov18} and Ly$\alpha$ escape at the systemic redshift \citep[e.g.][]{dijkstra16,matthee21,naidu22} are also considered strong indicators of LyC leakage, which is why double-peaked Ly$\alpha$ profiles are interesting for the study of LyC escape fractions. In addition, RXC0018-LAE1 appears to be a low-mass, irregular compact star-forming galaxy and thus likely represents an analog of high-redshift galaxies, in particular the faint low-mass population that drove the reionization of the Universe. These prospects, the double peak with a stronger blue component, together with the high magnification factors and the multiple imaging, make RXC0018-LAE1 particularly interesting for future studies to constrain LyC emission in the EoR.

This work is structured as follows: In section~\ref{sec:target} we present RXC0018-LAE1, data and observations. In section~\ref{sec:zsys} we first present near-infrared (NIR) spectroscopy targeting optical emission lines, which are key to the analysis of the double Ly$\alpha$ emission subsequently presented in section~\ref{sec:Lya}. In section~\ref{sec:SED-fitting} we present an SED fit to the galaxy. The results are then summarized in section~\ref{sec:conclusion}. Throughout this paper, we adopt a standard flat $\Lambda$CDM cosmology with $H_0=70\,\mathrm{km}\,\mathrm{s}^{-1}\,\mathrm{Mpc}^{-1}$, $\Omega_m=0.3$ and $\Omega_{\Lambda}=0.7$. All magnitudes are quoted in the AB system \citep{oke83}.

\section{Target and observations} \label{sec:target}

\begin{figure*}
    \centering
    \includegraphics[width=0.51\textwidth, keepaspectratio=true]{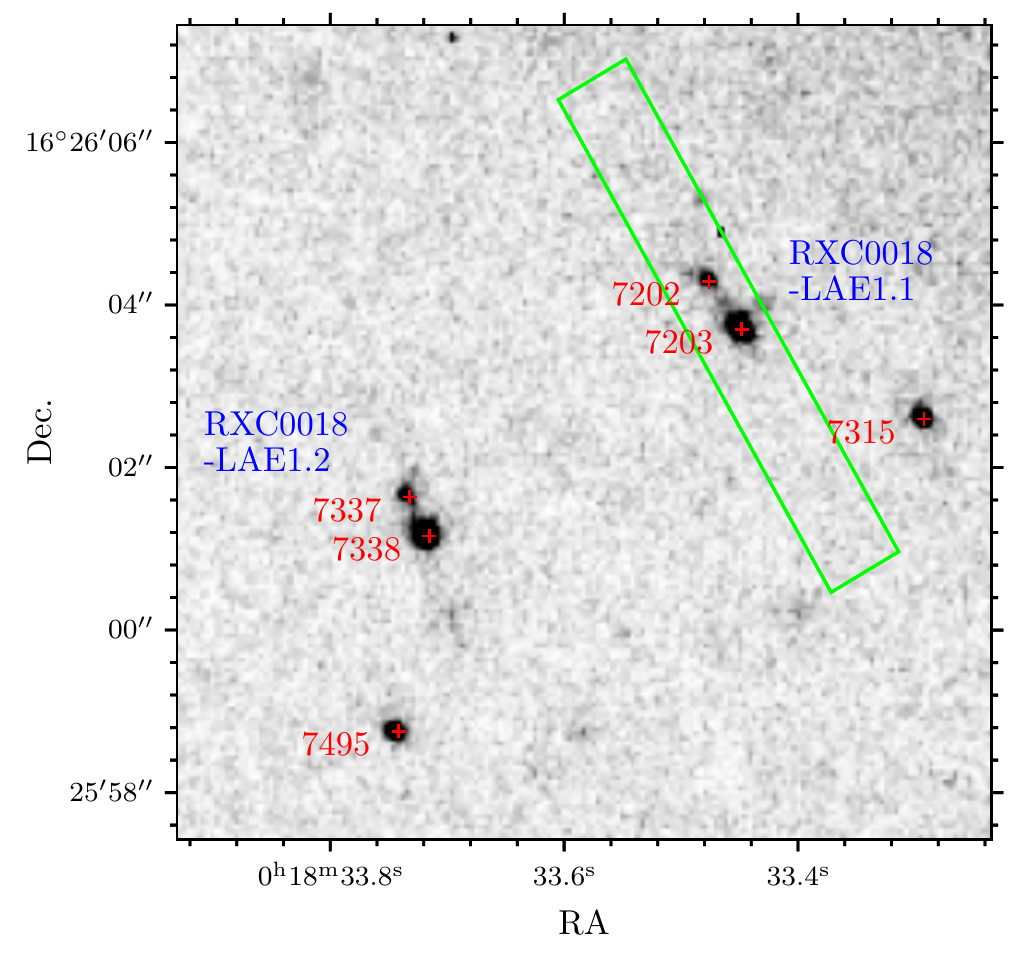}
    \includegraphics[width=0.47\textwidth, keepaspectratio=true]{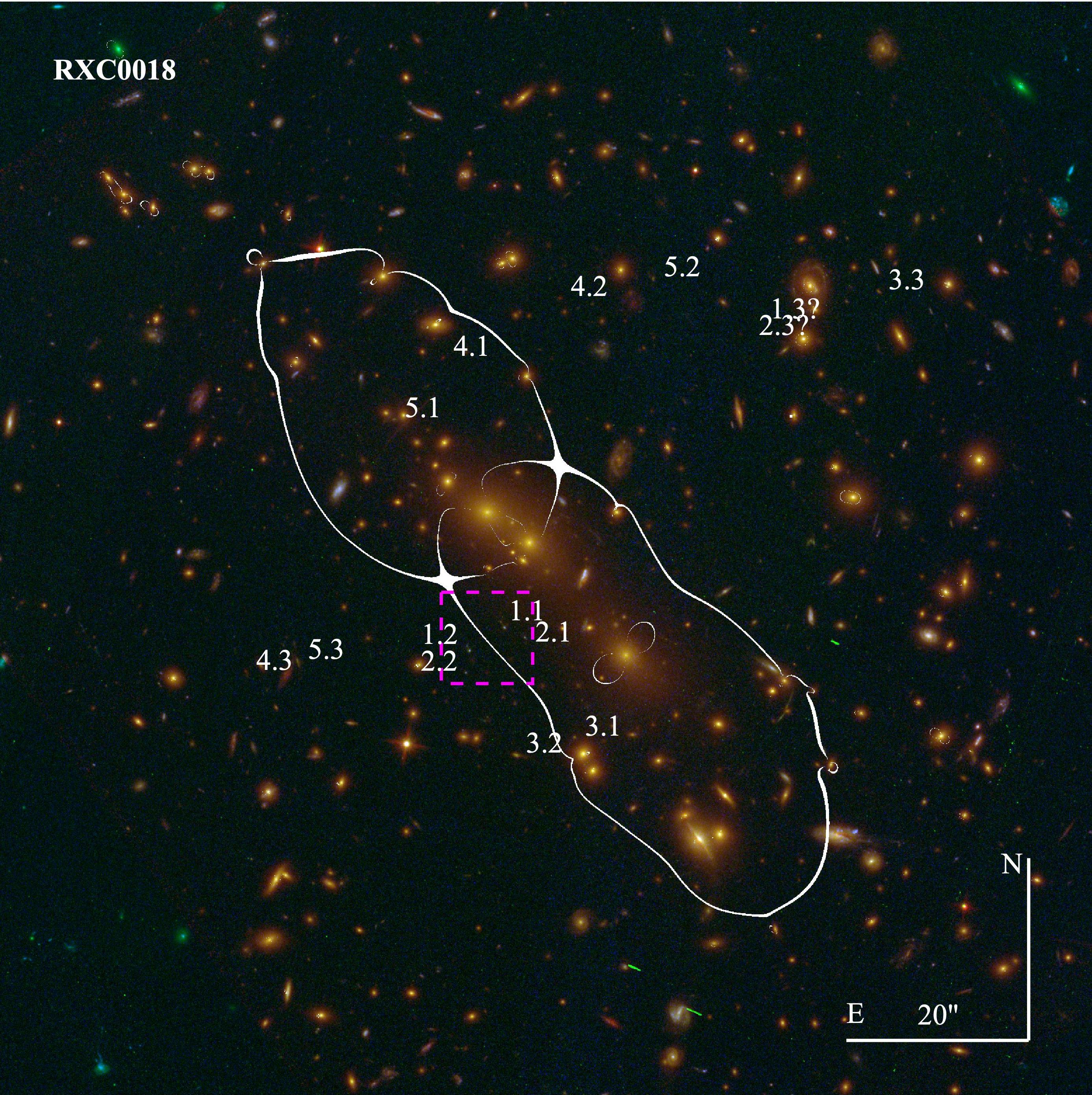}
    \caption{\textit{Left}: Center positions and RELICS catalog IDs (red) of each component of RXC0018-LAE1 over a stack of all 10 RELICS HST images of RXC0018 (see Tab.~\ref{tab:photometry}) in $0.06\arcsec$/pixel resolution. The green rectangle outlines the slit position and orientation used in our GNIRS observations presented in section~\ref{sec:spectroscopy}. The relative size of the MUSE detection can be seen in Fig.~\ref{fig:lya_S/N}. \textit{Right}: Color image of RXC0018 from the RELICS HST data. Numbered on the figure are multiple image families used as constraints for the modeling. Also shown in white are the critical curves for RXC0018-LAE1 (system 1 in the figure) at $z=3.218$. Additional counter images for systems 1 (RXC0018-7203 \& RXC0018-7338) and 2 (RXC0018-7315 \& RXC0018-7495) are expected on the other side of the cluster but not yet securely identified in the data. The magenta square represents the $10\arcsec\times10\arcsec$ cutout shown in the left-hand panel.}
    \label{fig:target}
\end{figure*}

\begin{table*}
    \centering
    \caption{Coordinates, photometric redshifts, gravitational-lensing magnifications and broad-band photometry of RXC0018-LAE1. The positions, photometric redshifts and HST photometry are taken from the public RELICS catalog available on \texttt{MAST} and presented in \citet{coe19}. The extraction of the \textit{K}-band photometry is detailed in section~\ref{sec:IR-photometry} and the magnifications are computed from our best-fitting lensing model presented in section~\ref{sec:SL}.}
    \begin{tabular}{lrrrrrr}
    \hline              &   \multicolumn{3}{c}{RXC0018-LAE1.1}                                                                                                                                  &   \multicolumn{3}{c}{RXC0018-LAE1.2}\\\hline
    RELICS ID           &   7203                                                &   7202                                                &   7315                                                &   7338                                                &   7337    &   7495\\\hline
    RA                  &   $0^{\mathrm{h}}18^{\mathrm{m}}33.450^{\mathrm{s}}$  &   $0^{\mathrm{h}}18^{\mathrm{m}}33.478^{\mathrm{s}}$  &   $0^{\mathrm{h}}18^{\mathrm{m}}33.294^{\mathrm{s}}$  &   $0^{\mathrm{h}}18^{\mathrm{m}}33.717^{\mathrm{s}}$  &   $0^{\mathrm{h}}18^{\mathrm{m}}33.734^{\mathrm{s}}$  &   $0^{\mathrm{h}}18^{\mathrm{m}}33.743^{\mathrm{s}}$\\
    Dec.                &   $16^{\circ}26\arcmin03.710\arcsec$                  &   $16^{\circ}26\arcmin04.297\arcsec$                  &   $16^{\circ}26\arcmin02.606\arcsec$                  &   $16^{\circ}26\arcmin01.165\arcsec$                  &   $16^{\circ}26\arcmin01.645\arcsec$  &   $16^{\circ}25\arcmin58.763\arcsec$\\
    $z_{\mathrm{phot}}$ &   $3.48_{-0.08}^{+0.07}$                              &   $3.40_{-3.30}^{+0.15}$                              &   $3.28_{-3.23}^{+0.12}$                              &   $3.42_{-0.09}^{+0.05}$                              &   $0.06_{-0.02}^{+3.25}$  &   $0.29_{-0.03}^{+3.26}$\\
    $\mu$               &   $6.5\pm0.2$                                         &   $6.6\pm0.2$                                         &   $4.2\pm0.1$                                         &   $10.1\pm0.7$                                        &   $9.7\pm0.6$             &   $6.4\pm0.4$\\\hline
    F435W               &   $25.64\pm0.11$                                      &   $26.70\pm0.24$                                      &   $25.93\pm0.18$                                      &   $25.50\pm0.08$                                      &   $26.33\pm0.14$          &   $26.43\pm0.15$\\
    F555W               &   $24.60\pm0.03$                                      &   $26.21\pm0.08$                                      &   $25.71\pm0.06$                                      &   $24.53\pm0.02$                                      &   $26.31\pm0.08$          &   $25.78\pm0.05$\\
    F606W               &   $24.60\pm0.03$                                      &   $26.32\pm0.18$                                      &   $25.48\pm0.04$                                      &   $24.55\pm0.03$                                      &   $26.05\pm0.08$          &   $25.76\pm0.06$\\
    F775W               &   $24.58\pm0.03$                                      &   $26.10\pm0.09$                                      &   $25.56\pm0.06$                                      &   $24.61\pm0.03$                                      &   $26.27\pm0.10$          &   $25.59\pm0.06$\\
    F814W               &   $24.59\pm0.03$                                      &   $26.24\pm0.08$                                      &   $25.57\pm0.04$                                      &   $24.56\pm0.02$                                      &   $26.20\pm0.07$          &   $25.69\pm0.04$\\
    F850LP              &   $24.40\pm0.06$                                      &   $26.10\pm0.17$                                      &   $25.83\pm0.15$                                      &   $24.73\pm0.07$                                      &   $26.52\pm0.23$          &   $25.43\pm0.10$\\
    F105W               &   $24.81\pm0.06$                                      &   $26.35\pm0.17$                                      &   $25.81\pm0.11$                                      &   $24.92\pm0.07$                                      &   $26.37\pm0.16$          &   $26.03\pm0.12$\\
    F125W               &   $24.94\pm0.13$                                      &   $26.06\pm0.23$                                      &   $25.89\pm0.21$                                      &   $25.16\pm0.14$                                      &   $27.37\pm0.59$          &   $26.06\pm0.22$\\
    F140W               &   $24.84\pm0.10$                                      &   $26.50\pm0.27$                                      &   $26.17\pm0.22$                                      &   $25.11\pm0.11$                                      &   $27.35\pm0.49$          &   $26.01\pm0.18$\\
    F160W               &   $24.94\pm0.07$                                      &   $26.31\pm0.16$                                      &   $26.11\pm0.14$                                      &   $25.01\pm0.07$                                      &   $26.44\pm0.17$          &   $26.02\pm0.12$\\
    \textit{Ks}         &   $24.88\pm0.22^{\mathrm{a}}$              &   $26.30\pm0.61^{\mathrm{a}}$                       &   $25.89\pm0.44^{\mathrm{a}}$                     &   $24.40\pm0.20$                                      &   $25.88\pm0.61$          &   $25.43\pm0.43$\\\hline
    \end{tabular}
    \par\smallskip
    $^{\mathrm{a}}$\, The \textit{Ks}-band photometry of RXC0018-LAE1.1 was not measured directly but inferred from the RXC0018-LAE1.2 fluxes using the magnification ratio.
    \label{tab:photometry}
\end{table*}

The object studied in this work, RXC0018-LAE1, is a strongly lensed system multiply imaged by the RELICS cluster RXC~J0018.5+1626 (RXC0018 hereafter). As can be seen in the left-hand panel of Fig.~\ref{fig:target}, each of the two counter images, designated as RXC0018-LAE1.1 and RXC0018-LAE1.2 respectively, is composed of one large and bright main component (RELICS IDs 7203 and 7338) and one smaller and fainter companion (RELICS IDs 7202 and 7337), separated by $0.71\arcsec$ ($0.54\arcsec$ in RXC0018-LAE1.2). According to our lensing model (see section~\ref{sec:SL}), this translates to a physical distance of $\simeq2.9$\,kpc in the source plane.

The HST imaging of RXC0018 comprises mosaics in 10 broad-band filters of the \textit{Advanced Camera for Survey} (ACS) and the \textit{Wide Field Camera Three} (WFC3): F435W, F555W, F606W, F775W, F814W, F850LP, F105W, F125W, F140W and F160W, which are publicly available in the RELICS repository on the \texttt{Mikulski Archive for Space Telescopes} (\texttt{MAST}). In this work we use the ACS+WFC3 catalog, also available in \texttt{MAST}, which contains HST fluxes obtained with \texttt{SExtractor} \citep{bertin96} and photometric redshifts computed with the \texttt{Bayesian Photometric Redshifts} code \citep[\texttt{BPZ};][]{benitez00,coe06}. We refer the reader to \citet{coe19} for the details of data reduction, catalog assembly and photometric redshift computation. The HST magnitudes and photometric redshifts used in our analysis are shown in Tab.~\ref{tab:photometry}.

The photometric redshift solution from the RELICS catalog places the main component of both images at roughly the same redshift $z_{\mathrm{phot}}\sim3.45$ within the uncertainties (see Tab.~\ref{tab:photometry}). In each image there is an additional third source (RELICS IDs 7315 and 7495, see Fig.~\ref{fig:target}) located $2.5\arcsec$ away from the main component ($2.4\arcsec$ in RXC0018-LAE1.2). While this source has a similar photometric redshift as the main component, the uncertainties make it unclear if this is a companion to the $z\simeq3.218$ source or if it lies at a somewhat different redshift. Note that if this source indeed lies at the same redshift as RXC0018-LAE1, it would only be $\simeq9.8$\,kpc away from the main component.

RXC0018 was also observed with the \textit{Multi Unit Spectroscopic Explorer} \citep[MUSE;][]{bacon10} on ESO's \textit{Very Large Telescope} (VLT) under program ID 0103.A-0777(B) (PI: A. Edge). The final reduced and calibrated data cube is publicly available on the ESO Science Archive. It achieves a $5\sigma$-depth of 23.13\,magnitudes, has a spatial pixel scale of 0.2\arcsec/pix, and a spectral resolution of $\Delta\lambda=1.25$\,\AA. Both RXC0018-LAE1.1 and RXC0018-LAE1.2 consistently show a clearly double-peaked strong emission line at $\sim5125$\,\AA\ which corresponds to Ly$\alpha$ at $z\simeq3.217$ \citep[based on the center between the two peaks; see][]{verhamme18} in the MUSE coverage of RXC0018. While one might consider the possibility that this double-peaked feature might be the \ion{Mg}{ii}]$\lambda\lambda2796$\AA,$2803$\AA~doublet at $z\simeq0.831$ or the [\ion{O}{ii}]$\lambda\lambda3726$\AA,$3729$\AA~doublet at $z\simeq0.375$, these possibilities are clearly ruled out by further spectroscopic analysis (see section~\ref{sec:zsys}) and the lensing configuration. We note that there also exist \textit{Magellan Clay Telescope}/\textit{Low Dispersion Survey Spectrograph} (LDSS3-C) observations of RXC0018-7203 taken on 2017, July 27 \citep[PI: K. Sharon; for more details see][]{mahler19} but the resolution is insufficient to resolve the double-peak. A detailed analysis of the double-peaked Ly$\alpha$ emission feature of RXC0018-LAE1 seen in the MUSE data is presented in section~\ref{sec:Lya}.

\subsection{Ancillary \textit{Ks}-band photometry} \label{sec:IR-photometry}
In addition to the RELICS photometry, we also make use of \textit{Ks}-band imaging data taken with the \textit{High Acuity Wide field K-band Imager} \citep[HAWK-I;][]{kissler-patig08} on VLT. The \textit{Ks}-band imaging of RXC0018 (Program ID: 0103.A-0871(B), PI: A. Edge) is publicly available on the ESO Science archive and achieves a $5\sigma$-depth of $26.15$\,magnitudes on a 0.11\arcsec/pix scale.

We use the \texttt{photutils} package \citep[\texttt{v1.3.0};][]{photutils21} to extract the \textit{K}-band photometry of our target. We subtract a 2D global background and then measure the flux in circular apertures of diameter $D_{\mathrm{ap}}=0.5\arcsec$ around our sources. The fluxes and their uncertainties are then corrected for aperture losses as $f_{\mathrm{tot}}=cf_{\mathrm{ap}}$ where $c=1.98$ is the aperture correction factor computed for a 0.5\arcsec aperture from the HAWK-I point-spread-function (PSF) in \citet{furtak21}. The resulting \textit{K}-band magnitudes for RXC0018-7338, RXC0018-7337 and RXC0018-7495 are shown in Tab.~\ref{tab:photometry}. We discard the \textit{K}-band photometry for the other image, RXC0018-LAE1.1, because we suspect it to be contaminated by some diffuse noise or possibly foreground source. This consistently also affects, e.g., the F850LP band in the HST imaging (Tab.~\ref{tab:photometry}) and the MUSE data (section~\ref{sec:flux_extraction}).

\subsection{Gemini GNIRS observations} \label{sec:spectroscopy}
In order to complement the existing optical MUSE spectroscopy and to precisely measure our target's systemic redshift, we observed RXC0018-7203 and RXC0018-7202 in the NIR range with the \textit{Gemini Near-Infrared Spectrograph} (GNIRS) on the Gemini-North telescope (Program ID: GN-2021B-Q-903; PI: A. Zitrin). Using the 110.5\,l/mm grating and the short blue camera in cross-dispersed mode and a slit of length 7\arcsec\ and width 1\arcsec, we obtained a total of 2\,h integration time on target. The slit position and orientation of this observation are also shown in Fig.~\ref{fig:target}.

The data were reduced, co-added, extracted and corrected for telluric line absorption with the Gemini IRAF package \texttt{v1.14}\footnote{\url{https://www.gemini.edu/observing/phase-iii/understanding-and-processing-data/data-processing-software/gemini-iraf-general}}. The orders of the cross-dispersed data were then separated into \textit{J}-, \textit{H}- and \textit{K}-band spectra respectively and flux-normalized to the broad-band photometry in each band given in Tab.~\ref{tab:photometry}. Because the GNIRS observations were targeted at RXC0018-LAE1.1 and we only have usable \textit{K}-band photometry for RXC0018-LAE1.2, as explained in section~\ref{sec:IR-photometry}, we use the ratio between the magnifications (see section~\ref{sec:SL} and Tab.~\ref{tab:photometry}) to convert the RXC0018-7338 \textit{K}-band flux to RXC0018-7203. Note that we also apply a heliocentric correction of $-11.3\,\frac{\mathrm{km}}{\mathrm{s}}$ to the spectra, to match the heliocentric-corrected MUSE data frame of reference. The final spectra have (observed) spectral resolutions of $\Delta\lambda=1.1$\,\AA, $\Delta\lambda=1.4$\,\AA\ and $\Delta\lambda=1.8$\,\AA\ respectively in the \textit{J}-, \textit{H}- and \textit{K}-bands.

\subsection{Lensing magnification} \label{sec:SL}
We construct a strong lensing (SL) model for RXC0018 using the \texttt{Light-Traces-Mass} method \citep[\texttt{LTM};][]{zitrin09a,zitrin15a}. We adopt five multiple image systems spanning a redshift range up to $z\sim5$. The resulting critical curves for our model are shown in the right-hand panel of Fig.~\ref{fig:target} together with the multiple image systems used as constraints. A first SL model for this cluster was published by \citet{zitrin11a} when analyzing the 12 $z>0.5$ MACS clusters \citep[][]{ebeling07}. Thanks to RELICS data we are able to identify several new sets of multiple image systems and measure photometric redshifts for them (in addition to RXC0018-LAE1 for which there is a spectroscopic measurements). We did not detect any emission lines for the other systems in the MUSE data cube. We use a positional uncertainty of 0.5\arcsec\ for multiple images in the modeling. The resulting best-fit model's image reproduction error is 0.86\arcsec. The SL magnifications for the sources analysed in this work are given in Tab.~\ref{tab:photometry}.

We note also that there is another parametric SL model for RXC0018 publicly available on the \texttt{MAST} archive, constructed with the SL modeling software \texttt{lenstool} \citep[][]{kneib96,jullo07,jullo09}. This model seems to predict somewhat higher magnifications at the coordinates of RXC0018-LAE1 than our \texttt{LTM} models (by factors up to $\sim2$). While we do not use this model here and defer a proper comparison of the lens models and their magnifications for future work, we note that higher magnifications would mean the absolute flux results of this work, i.e. integrated fluxes, luminosities and stellar mass, might be lower. 

\section{Systemic redshift from optical emission lines} \label{sec:zsys}

\begin{figure}
    \centering
    \includegraphics[width=\columnwidth, keepaspectratio=true]{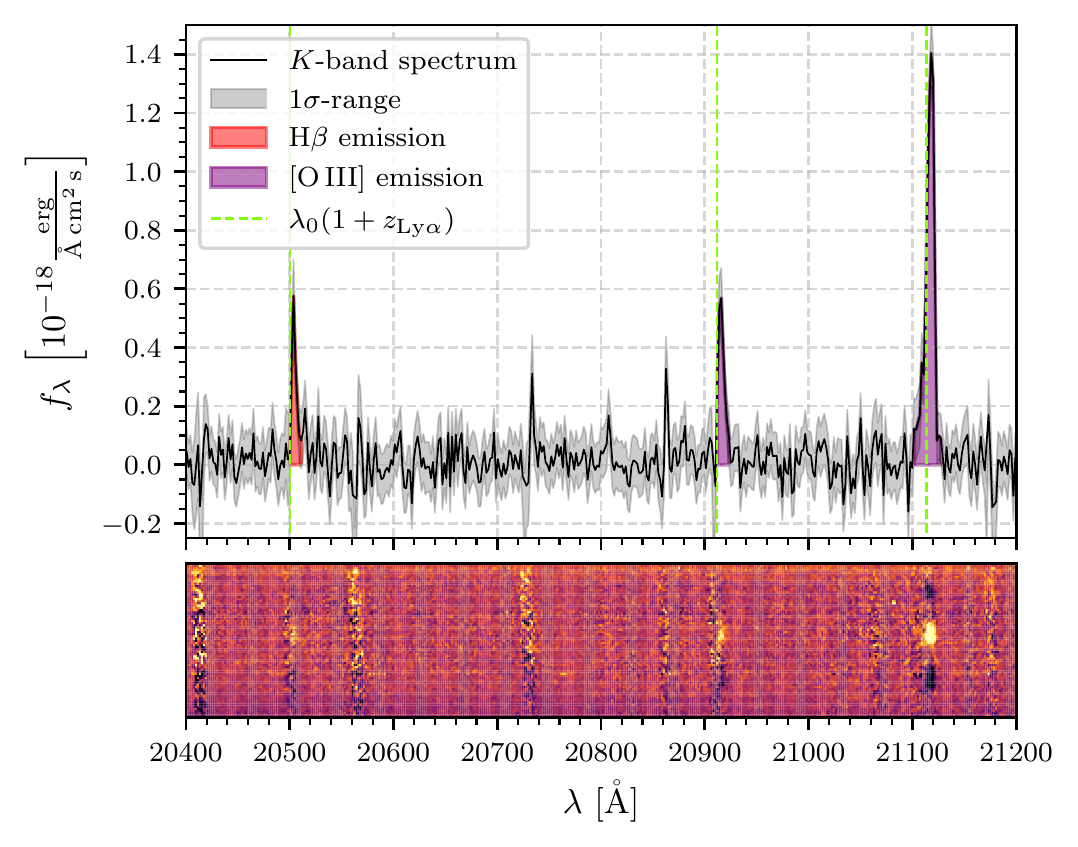}
    \caption{Observed GNIRS \textit{K}-band spectrum of RXC0018-7203 (black) and its $1\sigma$-range (grey shaded area). The three prominent emission features are identified as H$\beta$, [\ion{O}{iii}]$\lambda4959$\AA\ and [\ion{O}{iii}]$\lambda5007$\AA\ respectively. The red and purple shaded areas represent the integrated line fluxes and the green dashed lines show where the respective emission line can be expected according to the Ly$\alpha$ redshift derived from the MUSE data in section~\ref{sec:target}).}
    \label{fig:optical_emission_lines}
\end{figure}

\begin{table*}
    \centering
    \caption{GNIRS rest-frame optical spectroscopy results for RXC0018-7203. The integrated line fluxes were corrected for gravitational magnification given in Tab.~\ref{tab:photometry}.}
    \begin{tabular}{lcccccccc}
    \hline
    Line                                &   $\lambda_0$ [\AA]       &   $\lambda_{\mathrm{obs}}$ [\AA]  &   $z$     &   FWHM [\AA]                &   FWHM $\left[\frac{\mathrm{km}}{\mathrm{s}}\right]$ &   $F_{\mathrm{line}}$ $\left[10^{-18}~\frac{\mathrm{erg}}{\mathrm{s\,cm}^2}\right]$ &   EW$_{\mathrm{obs}}$ [\AA]   &   EW$_0$ [\AA]\\\hline
    H$\beta$                            &   4861.333$^{\mathrm{a}}$ &   $20504.0\pm0.8$                 &   $3.2178\pm0.0002$   & $6.3\pm1.4$   &   $92\pm30$   &   $0.42\pm0.06$                                       &   $140\pm49$                  &   $33\pm12$\\
    $[$\ion{O}{iii}$]\lambda4959$\AA    &   4958.911$^{\mathrm{a}}$ &   $20915.2\pm1.0$                 &   $3.2177\pm0.0002$   &   $9.0\pm1.5$ &   $130\pm32$  &   $0.56\pm0.05$                                       &   $187\pm62$                  &   $44\pm15$\\
    $[$\ion{O}{iii}$]\lambda5007$\AA    &   5006.843$^{\mathrm{a}}$ &   $21117.4\pm0.2$                 &   $3.2177\pm0.0001$   &   $8.0\pm0.6$ &   $114\pm15$  &   $2.00\pm0.10$                                       &   $672\pm217$                 &   $159\pm51$\\\hline
    \end{tabular}
    \par\smallskip
    $^{\mathrm{a}}$\,Taken from the \texttt{NIST} Atomic Spectra Database, available at \url{https://physics.nist.gov/asd}. Note that quoted here are air wavelengths.
    \label{tab:line-fits}
\end{table*}

Our GNIRS observations yielded \textit{J}-, \textit{H}- and \textit{K}-band spectra for RXC0018-7203 and RXC0018-7202 covering wavelengths from 12055\,\AA\ to 13230\,\AA, 15070\,\AA\ to 16480\,\AA, and 20085\,\AA\ to 21965\,\AA, respectively. While we do not find any significant emission in the \textit{J}- and \textit{H}-bands, our \textit{K}-band spectrum shows three strong emission features as can be seen in Fig.~\ref{fig:optical_emission_lines}. These are consistent with the optical [\ion{O}{iii}] doublet, [\ion{O}{iii}]$\lambda5007$\AA\ and [\ion{O}{iii}]$\lambda4959$\AA, and the H$\beta$-line at the redshift expected from the MUSE and photometric data (see section~\ref{sec:target}). We use the \texttt{specutils} package \citep[\texttt{v1.5.0};][]{specutils21} to perform a Gaussian fit to each emission line and thus obtain line centroids and widths. In order to accurately propagate the pixel-wise variance of the data to the line parameters, we evaluate the joint posterior distribution of 20 \textit{Monte-Carlo Markov Chains} (MCMC) of $10^4$ steps each, run with the \texttt{emcee} package \citep{foreman-mackey13}, and compute the uncertainties of each fit-parameter. We integrate the observed spectrum at each emission line to obtain line fluxes as shown by the shaded areas in Fig.~\ref{fig:optical_emission_lines} and correct them for magnification. The wavelength window for the flux integration is chosen using the $99.7$\,\%-range ($3\sigma$) of the Gaussian fit to each line. All measured properties of the emission lines are summed up in Tab.~\ref{tab:line-fits}.

Using the centroids of the Gaussian fits to the three optical emission lines and the associated uncertainties, we find a weighted average systemic redshift of $z_{\mathrm{sys}}=3.2177\pm0.0001$ for RXC0018-7203. These clear detections of the rest-frame optical lines in the \textit{K}-band spectrum further rule out the lower redshift \ion{Mg}{ii}] or [\ion{O}{ii}] doublet scenarios for the double-peaked emission line in the MUSE spectra and firmly confirm it to indeed be the Ly$\alpha$ line. Note that we do not observe any significant spatial offset that could correspond to the separation of RXC0018-7203 and RXC0018-7202 in the HST imaging (see Fig.~\ref{fig:target} and section~\ref{sec:target}). The separation of 0.71\arcsec is however at the limit of the spatial resolution obtained in our GNIRS (0.15\arcsec/pixel) observations given the average seeing of $\sim0.7\arcsec$ (varying between $\sim0.5\arcsec$ and $\sim1\arcsec$ over the 2\,h of observations). We therefore conclude that RXC0018-7202 either does not contribute significant flux to the measured emission lines, which is plausible given its faintness ($\sim26$\,magnitudes) in the \textit{K}-band, or it is indistinguishably blended into the line emission of RXC0018-7203.

We compute the optical line EWs using the \textit{K}-band flux shown in Tab.~\ref{tab:photometry} and taking the magnification ratio into account as explained in section~\ref{sec:spectroscopy}. As can be seen in Tab.~\ref{tab:line-fits}, we find relatively low optical EWs of order $\mathrm{EW}_0\sim10-100$\,\AA\ in RXC0018-7203 which sets RXC0018-LAE1 slightly apart from objects typically used as high-redshift analogs, such as e.g. extreme emission line galaxies (EELGs) with [\ion{O}{iii}] EWs $\gtrsim1000$\,\AA\ \citep[e.g.][]{atek11,atek14b,maseda14,maseda18,reddy18b,jaskot19,tang19,boyett22} or `Green Pea' galaxies which also typically display very strong optical emission lines \citep[e.g.][]{cardamone09,izotov11,amorin12,jaskot13,henry15}. Optical emission lines with low EWs on the other hand could however be an indication of LyC escape as was found by \citet{zackrisson13,zackrisson17}: According to these works, $\mathrm{EW}_{\mathrm{H}\beta,0}\lesssim30$\,\AA\ in particular would imply LyC escape fractions $f_{\mathrm{esc,LyC}}\gtrsim0.5$.

We also find an [\ion{O}{iii}]$\lambda$5007\AA/H$\beta$ ratio of $4.8\pm0.7$ which is relatively high for a star-forming galaxy but slightly lower than the ratios typically found in EELGs \citep[e.g.][]{tang19}. We do not detect any emission from the [\ion{O}{ii}]$\lambda\lambda3726$\AA,$3729$\AA~doublet in the \textit{H}-band spectrum with a magnification-corrected $3\sigma$ upper limit of $<0.22\times10^{-18}\frac{\mathrm{erg}}{\mathrm{s\,cm}^2}$ (see appendix~\ref{app:OII_upper_limit}). This results in a $3\sigma$ lower limit on the [\ion{O}{iii}]/[\ion{O}{ii}]$\lambda3726$\AA$+3729$\AA~ratio of $>9.3$. Note that high [\ion{O}{iii}]/[\ion{O}{ii}] ratios have been found to correlate with LyC leakage \citep{izotov16,izotov17,izotov18,fletcher19,jaskot19,flury22b} but do not necessarily represent a sufficient condition for high LyC escape fractions \citep{stasinska15,shapley16,naidu18,izotov18,bassett19,jaskot19,tang19,plat19}.

\section{Double-peaked Lyman alpha} \label{sec:Lya}

\begin{figure}
    \centering
    \includegraphics[width=\columnwidth, keepaspectratio=true]{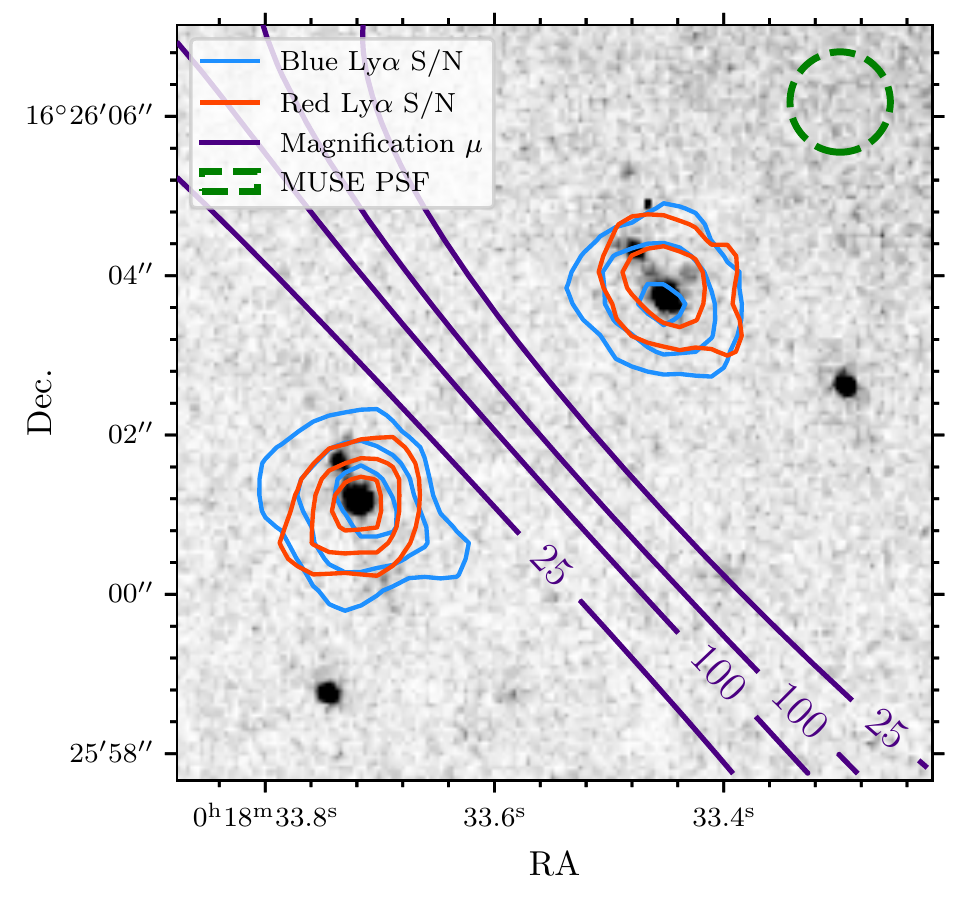}
    \caption{MUSE S/N-contours at $\lambda=5122.6$\,\AA\ (blue) and $\lambda=5128.9$\,\AA\ (red), corresponding to the two Ly$\alpha$ peaks of RXC0018-LAE1 over the same RELICS HST image stack as in Fig.~\ref{fig:target}. The blue contours are computed from a stack of 4 elements on the spectral axis and the red contours from a stack of 3 spectral elements around the central wavelength. The contours represent, from outer to inner, $2\sigma$, $3\sigma$ and $4\sigma$. Note that in both images both Ly$\alpha$ peaks are centered on the main component of the HST imaging. We show gravitational magnification contours for a source at $z=3.218$ according to our \texttt{LTM} model (see section~\ref{sec:SL}) as dark blue lines. The critical line passes right between the two images RXC0018-LAE1.1 and RXC0018-LAE1.2. The green circle represents the FWHM of the MUSE PSF (1.25\arcsec).}
    \label{fig:lya_S/N}
\end{figure}

With the systemic redshift in hand, we can confirm that the double-peaked emission feature detected in the MUSE data is indeed the Ly$\alpha$ line at $z=3.218$. Note that we also tentatively detect other rest-frame UV emission lines in the MUSE data of RXC0018-LAE1 as detailed in appendix~\ref{app:UV_lines}. These are however relatively weak and we therefore defer more detailed analysis to future work.

Fig.~\ref{fig:lya_S/N} shows MUSE signal-to-noise contours of the two Ly$\alpha$ components. In both images the Ly$\alpha$ emission is consistently centered on the main component, i.e. RXC0018-7203 and RXC0018-7338 respectively. This rules out the possibility of the double peak arising from a separate Ly$\alpha$ emission from the secondary component at a slightly different redshift and instead confirms that the double peak is therefore genuine and originates from the same region in space. The figure also clearly shows that both the blue and the red Ly$\alpha$ emission have fairly similar and symmetric morphologies. Since gravitational lensing is achromatic, this means both peaks are equally affected by possible differential magnification due to the proximity with the critical line which passes in between the two images, RXC0018-LAE1.1 and RXC0018-LAE1.2 as is also shown in Fig.~\ref{fig:lya_S/N}. We can therefore safely ignore differential magnification effects in the analysis of this galaxy's Ly$\alpha$ profile.

Recent studies have revealed that LAEs commonly have Ly$\alpha$ halos of relatively low surface brightness which extend far beyond, up to $10\times$, the spatial extent of the stellar continuum emission detected in broad-band photometry \citep[e.g.][]{hayashino04,steidel11,momose14,momose16,wisotzki16,leclercq17,cai18,wisotzki18,cantalupo19,leclercq20}. Since the halo typically represents $\sim65\,\%$ of the total Ly$\alpha$ emission \citep{leclercq17}, we need to take special care in the extraction to include \textit{all} of the Ly$\alpha$ flux as detailed in the next section.

\subsection{Ly$\alpha$ flux extraction} \label{sec:flux_extraction}

\begin{figure}
    \centering
    \includegraphics[width=\columnwidth, keepaspectratio=true]{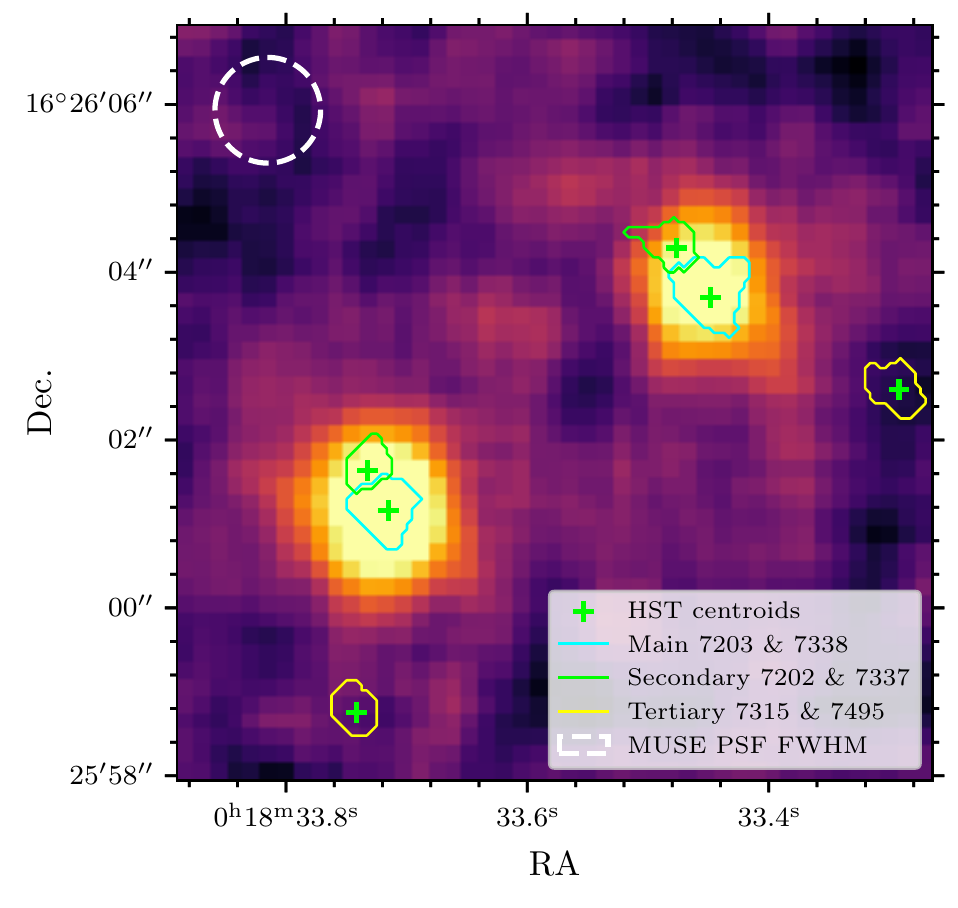}
    \caption{Continuum-subtracted full Ly$\alpha$ (15\,\AA) NB image of RXC0018-LAE1 smoothed with a 2D Gaussian kernel of $\sigma=0.5\arcsec$. The green crosses represent the coordinates of the RELICS HST detections and the white dashed circle the FWHM of the MUSE PSF. The solid contours represent the segmentations of the RELICS HST detections.}
    \label{fig:MUSE_NB_image}
\end{figure}

In order to extract all the Ly$\alpha$ flux of our sources and account for the emission in the low-surface-brightness halo, we follow the approach generally established for this kind of study with MUSE data \citep[e.g.][]{wisotzki16,leclercq17,drake17a,drake17b,hashimoto17,kerutt22}. We first use the \texttt{spectral-cube} package \citep[\texttt{v0.2};][]{spectral-cube14} to cut out a $16\arcsec\times16\arcsec$ region that encompasses both images of RXC0018-LAE1 on the two spatial axes and then use a median filter of 200 pixels on the spectral axis (250\,\AA) to construct a continuum-only cube. The spectral window used for the filter has been found to effectively remove emission lines in MUSE data \citep[][]{herenz17a}. We then subtract the continuum cube from the original $16\arcsec\times16\arcsec$ cube and construct three narrow-band (NB) images: A total NB image encompassing the total Ly$\alpha$ emission, one centered on the blue peak and one on the red peak. Note that we find residual continuum at the position of RXC0018-LAE1.1 in the continuum-subtracted cube. We therefore adopt the flux measurements of the counter image RXC0018-LAE1.2 as fiducial results in the following.

The NB images have bandwidths of 15\,\AA\ for the total Ly$\alpha$ NB image and 7\,\AA\ and 8\,\AA\ each for the blue and red NB images respectively. These were chosen as the limits where the continuum-subtracted MUSE flux density crosses zero on either side of the double-peaked Ly$\alpha$ emission. The total Ly$\alpha$ NB image of RXC0018-LAE1 is shown in Fig.~\ref{fig:MUSE_NB_image}. We use \texttt{photutils} (see section~\ref{sec:IR-photometry}) to measure the centroids and kron radii \citep{kron80} of the Ly$\alpha$ emission on both multiple images. Consistently in both RXC0018-LAE1.1 and RXC0018-LAE1.2, the centroids of the blue and the red peak align with the total Ly$\alpha$ peak and the RELICS catalog coordinates within $\lesssim0.1\arcsec$. This further confirms that the total Ly$\alpha$ emission of RXC0018-LAE1 originates from its main component and thus that the double peak is genuine.

Finally, continuum-subtracted spectra are extracted in a circular aperture of 3 Kron radii around the total Ly$\alpha$ centroids and collapsed to 1D-spectra. The chosen aperture size has been found to reliably extract Ly$\alpha$ halo fluxes as discussed extensively in \citet{kerutt22}. We integrate the obtained spectra in the bandwidth of the total Ly$\alpha$ NB image to obtain total Ly$\alpha$ fluxes for both images.

\subsection{UV continuum} \label{sec:UV_continuum}

\begin{figure}
    \centering
    \includegraphics[width=\columnwidth, keepaspectratio=true]{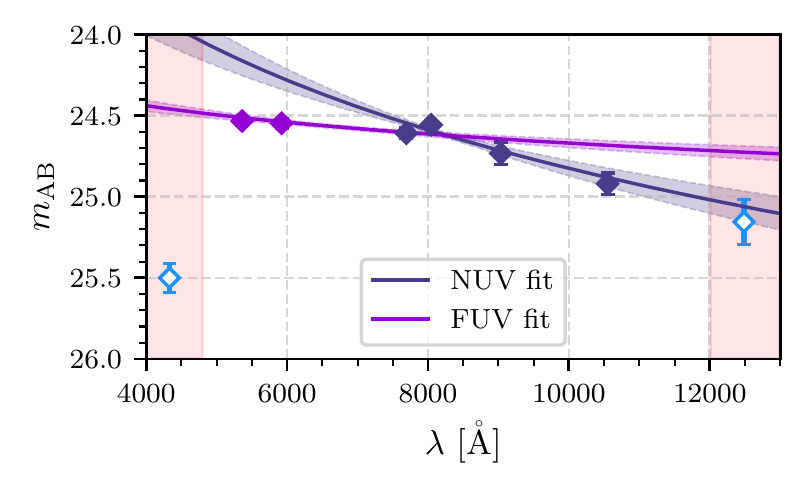}
    \caption{Power-law fits to the rest-frame UV photometry of RXC0018-7883. The dark purple line represents our NUV fit ($\lambda_0=1700-2400$\,\AA, dark purple points) and the light purple line the FUV fit to the same bands in addition to the two bluer filters (light purple points). The adjacent filters, not used in either fit are shown as open blue points. The red shaded areas delimit the full $\lambda_0=1100-2400$\,\AA\ range used in the FUV fit.}
    \label{fig:UV-continuum}
\end{figure}

In order to compute the Ly$\alpha$ EW of RXC0018-LAE1, we measure the continuum following the method detailed in \citet{hashimoto17} and fit a power-law relation

\begin{equation} \label{eq:UV-slope}
    m(\lambda)=-2.5\log(\lambda^{\beta+2})+m_0
\end{equation}

where $\beta$ is the UV-continuum slope and $m_0$ a constant to the rest-frame UV photometry. We fit this relation for the main component of RXC0018-LAE1.2, RXC0018-7338, because of the possible foreground contamination in RXC0018-LAE1.1 mentioned in sections~\ref{sec:IR-photometry} and~\ref{sec:flux_extraction}. Because there appears to be a flattening of the UV continuum towards the two bluer filters (see Tab.~\ref{tab:photometry}), we fit the continuum twice: One near UV (NUV) continuum spanning a rest-frame wavelength range from $1700$\,\AA\ to $2400$\,\AA\ (filters F775W, F815W, F850LP and F105W) as in e.g. \citet{bouwens09b,hashimoto17}, and a far UV (FUV) continuum including the two bluer filters F606W and F555W, i.e. down to rest-frame 1100\,\AA. The fits are performed with 20 MCMC chains of $10^4$ steps each. Both fits are shown in Fig.~\ref{fig:UV-continuum}. The resulting UV-slopes are both very blue which indicates a very low dust attenuation and might also indicate possible LyC escape \citep[e.g.][also see discussion in section~\ref{sec:SED-fitting}]{zackrisson13,zackrisson17}. Interestingly, we find a marked difference between the FUV and NUV slopes $\beta_{\mathrm{FUV}}=-2.23\pm0.06$ and $\beta_{\mathrm{NUV}}=-3.0\pm0.2$.

Using this UV-continuum fit, we infer the flux density at rest-frame $1500$\,\AA\ and from that compute the absolute UV magnitude of the main component of RXC0018-LAE1 which yields $M_{\mathrm{UV}}=-18.58\pm0.07$.

\subsection{Ly$\alpha$ EW and emission profile} \label{sec:Lya_properties}

\begin{figure}
    \centering
    \includegraphics[width=\columnwidth]{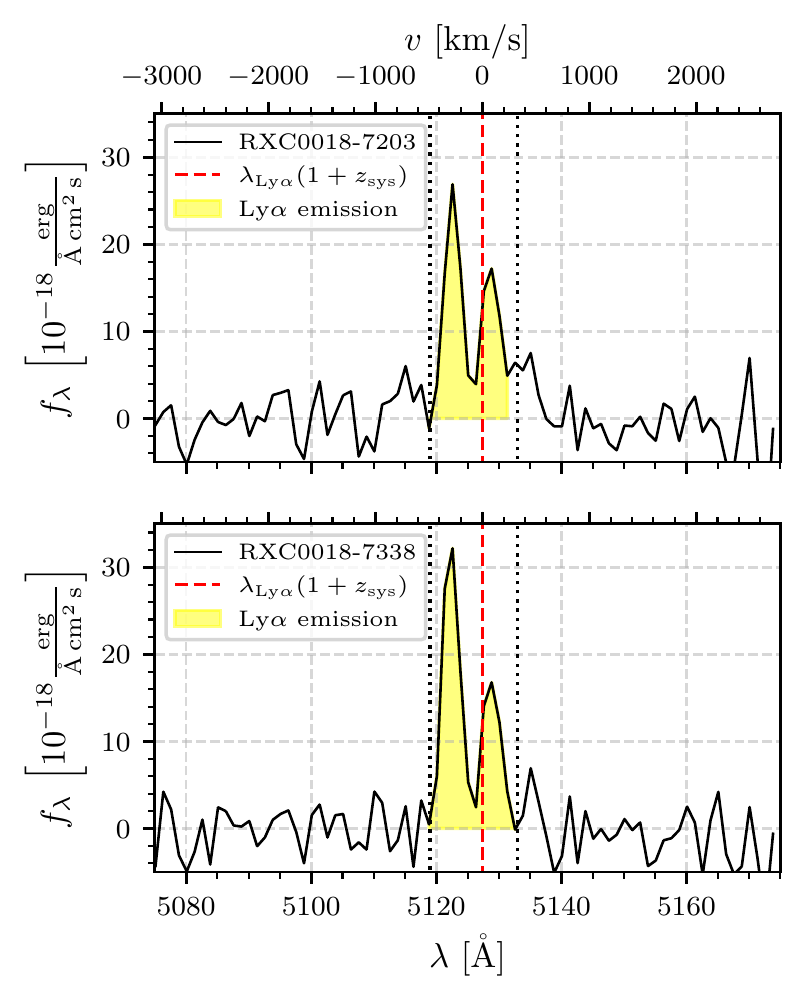}
    \caption{Continuum-subtracted Ly$\alpha$ emission in RXC0018-LAE1.1 (upper panel) and its counter image RXC0018-LAE1.2 (lower panel) with the systemic redshift marked as the red dashed line. The uncertainty of the systemic redshift is too small to be visible in the figure. Unlike other double-peaked LAEs observed to date, the blue peak is significantly stronger than the red peak and the two peaks show asymmetric shifts in velocity space. The yellow shaded area represents the total integrated Ly$\alpha$ flux and the black dotted lines the bandwidth of the Ly$\alpha$ NB image (see section~\ref{sec:flux_extraction}).}
    \label{fig:lya_emission}
\end{figure}

\begin{figure}
    \centering
    \includegraphics[width=\columnwidth, keepaspectratio=true]{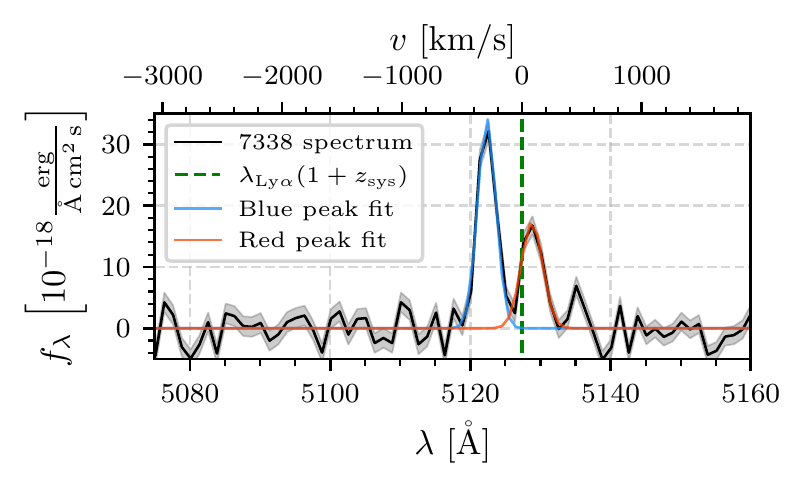}
    \caption{Continuum-subtracted Ly$\alpha$ emission of RXC0018-LAE1.2 with its $1\sigma$-range (grey shaded area) and the systemic redshift marked as the green dashed line. The two Gaussian fits to the two Ly$\alpha$ peaks are shown as blue and red lines respectively.}
    \label{fig:lya_emission_fits}
\end{figure}

\begin{figure}
    \centering
    \includegraphics[width=\columnwidth, keepaspectratio=true]{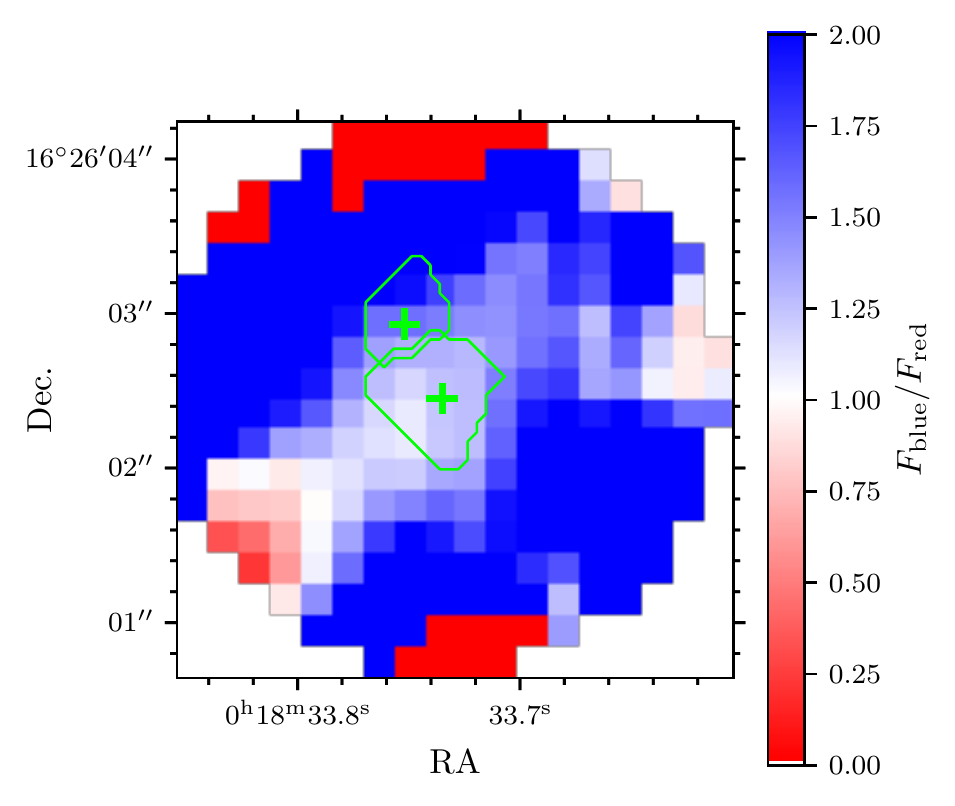}
    \caption{Blue-to-red peak ratio map for RXC0018-LAE1.2. The figure shows the 3 Kron radii area extracted from the continuum-subtracted MUSE cube (see section~\ref{sec:flux_extraction}). The green crosses and curves show the HST continuum centroids and segmentations respectively, like in Fig.~\ref{fig:MUSE_NB_image}.}
    \label{fig:peak-ratio}
\end{figure}

We measure a total magnification-corrected integrated Ly$\alpha$ flux of $F_{\mathrm{Ly}\alpha}=(17\pm1)\times10^{-18}\,\frac{\mathrm{erg}}{\mathrm{s}\,\mathrm{cm}^2}$ which yields a relatively high Ly$\alpha$ luminosity of $L_{\mathrm{Ly}\alpha}=(6.7\pm0.5)\times10^{42}\,\frac{\mathrm{erg}}{\mathrm{s}}$ for RXC0018-LAE1. Using the rest-frame UV-continuum fit obtained in section~\ref{sec:UV_continuum} to infer the continuum at Ly$\alpha$, this results in a rest-frame EW of $\mathrm{EW}_0=(63\pm2)$\,\AA.

The continuum-subtracted 1D-spectra of the two multiple images are shown in Fig.~\ref{fig:lya_emission}. They clearly display the double-peaked nature of the Ly$\alpha$ emission of RXC0018-LAE1. We fit the two peaks to obtain line centers and widths with the same method as for the optical emission lines in section~\ref{sec:zsys} and find line widths of $\mathrm{FWHM}_{\mathrm{blue}}=(179\pm37)\,\frac{\mathrm{km}}{\mathrm{s}}$ and $\mathrm{FWHM}_{\mathrm{red}}=(211\pm41)\,\frac{\mathrm{km}}{\mathrm{s}}$. The two Gaussian fits for RXC0018-7338 are shown in Fig.~\ref{fig:lya_emission_fits}. We also note a pronounced asymmetry in velocity space with regard to the systemic redshift: While the two peaks are separated by $\Delta v=(376\pm32)\,\frac{\mathrm{km}}{\mathrm{s}}$, the red peak lies much closer, $\Delta v_{\mathrm{red}}=(83\pm11)\,\frac{\mathrm{km}}{\mathrm{s}}$, to the systemic redshift than the blue peak which is shifted much further by $\Delta v_{\mathrm{blue}}=(-293\pm10)\,\frac{\mathrm{km}}{\mathrm{s}}$. Recently, \citet{verhamme18} measured a correlation between the peak separation $\Delta v$, red peak shift $\Delta v_{\mathrm{red}}$ from the systemic velocity and the (red-peak) Ly$\alpha$ FWHM in double-peaked LAEs. RXC0018-LAE1 significantly deviates (by $>2\sigma$) from the first of these relations with $\Delta v_{\mathrm{red}}(\Delta v)=(185\pm43)\,\frac{\mathrm{km}}{\mathrm{s}}$ inferred by the peak separation. The second one however, $\Delta v_{\mathrm{red}}(\mathrm{FWHM}_{\mathrm{red}})=(156\pm77)\,\frac{\mathrm{km}}{\mathrm{s}}$ computed using the FWHM of the red peak, agrees by $\sim1\sigma$ with our measured red peak shift from the systemic velocity. Note that the uncertainties on all of our velocity measurements are relatively large due to the low spectral resolution of MUSE which might in part explain these deviations.

The most prominent feature of this Ly$\alpha$ emission profile is that the blue peak is significantly stronger than the red peak with a blue-to-red peak ratio of $1.7\pm0.1$ ($\mathrm{EW}_{\mathrm{blue},0}=40\pm2$\,\AA\ and $\mathrm{EW}_{\mathrm{red},0}=24\pm2$\,\AA). We show the spatial distribution of the peak ratios for RXC0018-LAE1.2 in Fig.~\ref{fig:peak-ratio}, similarly to \citet{erb18}. As could already be expected from the signal-to-noise curves shown in Fig.~\ref{fig:lya_S/N}, the whole Ly$\alpha$ emission is dominated by the blue peak and the peak ratio becomes closer to unity towards the centroid of the emission where the slightly more compact red peak is also strongest. This is similar to what \citet{erb18} found, i.e. that the blue component becomes stronger towards the outer regions, even though it overall dominates the whole Ly$\alpha$ emission region in the case of RXC0018-LAE1. This blue-to-red peak ratio places RXC0018-LAE1 in stark contrast to other double-peaked LAEs observed to date. Indeed, the vast majority of double-peaked LAEs in large samples are dominated by their red peak \citep[e.g.][]{leclercq17,erb18,izotov18,matthee21,kerutt22,naidu22}. In some rare individual cases both peaks are almost equal \citep[e.g.][]{trainor15,leclercq17,izotov18,naidu22}, roughly equal but extremely faint \citep[e.g.][]{erb14}, or a slightly stronger blue peak is located red-ward of the systemic redshift \citep{endsley22}. Note that several instances of double-peaked Ly$\alpha$ with stronger blue peaks have been detected in extended Ly$\alpha$ blobs \citep{vanzella17,erb18,ao20,li22a}. In these cases the two Ly$\alpha$ peaks usually originate from distinct spatially separated objects or regions within the extended Ly$\alpha$ blob however whereas both the blue and the red peak clearly originate from the same galaxy in RXC0018-LAE1 (see Fig.~\ref{fig:lya_S/N}).

A dominant red peak is usually interpreted as evidence that the Ly$\alpha$ photons scatter through an outflowing medium \citep[e.g.][]{dijkstra06,verhamme06,barnes11,yang14,gronke16,gurung-lopez19}. The blue peak being stronger on the other hand could indicate an inflow of circumgalactic gas into the Ly$\alpha$ emitting galaxy \citep[e.g.][]{dijkstra06,verhamme06,yang14}. Since gas accretion has been found to preferentially occur along filamentary streams \citep[e.g.][]{keres05,keres09,dekel09}, this could potentially imply low covering fractions of neutral hydrogen and thus low \ion{H}{i} column density channels which could also allow LyC photons to escape. This however also strongly depends on other parameters such as velocity and gas temperature \citep[e.g.][]{kakiichi21,li22b}. Another indication of possible LyC escape would be a detection of Ly$\alpha$ photons escaping in the valley between the peaks near the line center \citep[e.g.][]{dijkstra16,vanzella16,rivera-thorsen17,gazagnes20,matthee21,naidu22}. While our spectra in Fig.~\ref{fig:lya_emission} hint at a possible $f_{\lambda}>0$ between the two peaks, the spectral resolution of the MUSE data is not sufficient to properly resolve this. It will therefore require further spectroscopic observations with very high spectral resolution to determine if this galaxy has Ly$\alpha$ photons escaping at the line center.

\section{Galaxy properties} \label{sec:SED-fitting}

\begin{figure}
    \centering
    \includegraphics[width=\columnwidth, keepaspectratio=true]{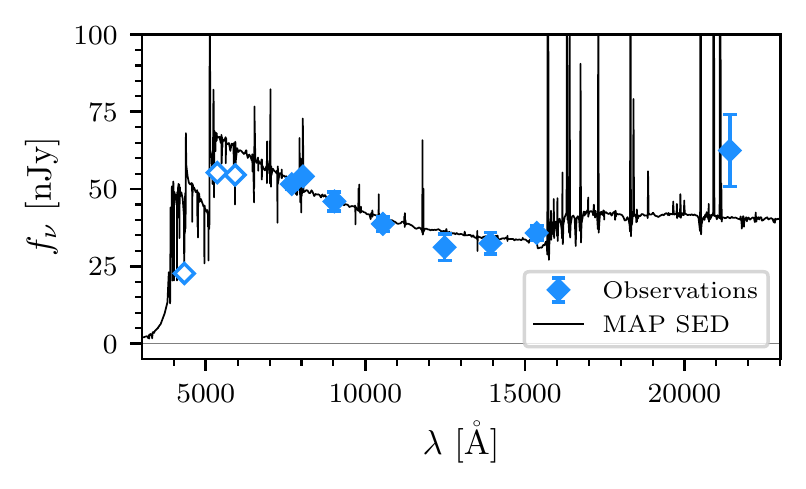}
    \caption{Best-fit \texttt{BEAGLE} SED for RXC0018-LAE1. The magnification-corrected photometry is shown as blue diamonds. Solid data points represent filters used in the fit and open points the filters excluded. The black curve represents the maximum-a-posteriori spectrum predicted by \texttt{BEAGLE}.}
    \label{fig:SED-fit}
\end{figure}

In order to place these Ly$\alpha$ results into context, we infer galaxy parameters by fitting spectral energy distributions (SED) to the broad-band photometry and emission line EWs with the \texttt{BayEsian Analysis of Galaxy sEds} tool \citep[\texttt{BEAGLE};][]{chevallard16} which is optimized to probe a large parameter space using a bayesian MCMC approach. We use stellar and nebular SED templates by \citet{plat19}, which combine the latest version of the stellar population templates by \citet{bc03} with the photoionization code \texttt{CLOUDY} \citep{ferland13}, and account for IGM attenuation with absorption models by \citet{inoue14}. Note that with the \citet{plat19} templates, the fraction $f_{\mathrm{esc},\mathrm{LyC}}$ of LyC photons escaping from density-bounded \ion{H}{ii} regions is a free parameter in our SED-fit. This is important because $f_{\mathrm{esc},\mathrm{LyC}}>0$ can significantly affect the galaxy SED, weakening both nebular lines and continuum emission \citep[e.g.][]{zackrisson13,zackrisson17}. The latter in particular also tends to make the galaxy SED bluer in the UV which might account for the extremely steep UV-slopes that we find (see section~\ref{sec:UV_continuum}).

The fit is performed for the main component of RXC0018-LAE1 for which we use the magnification-corrected photometry of RXC0018-7338 and the optical EWs measured in section~\ref{sec:zsys}. Since neither Ly$\alpha$ nor IGM radiative transfer are straight-forward to model, we exclude the measured Ly$\alpha$ EW and the three bluest filters (F435W, F555W and F606W) from the fit. We assume a delayed star-forming history (SFH), which goes as $\psi(t)\propto t\exp(-t/\tau)$, with the possibility for an ongoing star-burst for the last $10$\,Myr and model the dust attenuation using an SMC-like dust extinction law \citep[][]{pei92} which has been found to well fit star-forming galaxies at $z\sim3$ \citep[e.g.][]{reddy18a}, in particular at low metallicities \citep{shivaei20}. The best-fit maximum-a-posteriori (MAP) SED is shown in Fig.~\ref{fig:SED-fit} and the fit results in a low stellar mass of $\log(M_{\star}/\mathrm{M}_{\odot})=7.50_{-0.21}^{+0.34}$, a stellar age of $\log(t_{\mathrm{age}}/\mathrm{yr})=8.66_{-0.48}^{+0.37}$ and a moderate current star-formation rate (SFR) $\log(\psi/\mathrm{M}_{\odot}\,\mathrm{yr}^{-1})=0.18_{-0.02}^{+0.02}$. These results again place RXC0018-LAE1 apart from the EELGs typically considered as high-redshift analogs (see section~\ref{sec:zsys}) with very young ages and extreme ongoing star-formation activity \citep[e.g.][]{atek11,atek14b,maseda14,maseda18,reddy18b,tang19,boyett22}. The low metallicity and dust attenuation optical depth that we find, $\log(Z/\mathrm{Z}_{\odot})=-0.99_{-0.13}^{+0.24}$ and $\hat{\tau}_V=0.009_{-0.006}^{+0.010}$, are however in line with what can be expected for a high-redshift analog. The posterior distribution of our SED-fit predicts relatively high LyC escape fractions of $f_{\mathrm{esc},\mathrm{LyC}}>0.7$ at $3\sigma$ because these models provide the best interpretation of the low optical emission line EWs and the extremely blue UV-slopes that we measure (see sections~\ref{sec:zsys} and~\ref{sec:UV_continuum}). Note that the galaxy parameters inferred above and the optical emission features presented in section~\ref{sec:zsys} make RXC0018-LAE1 similar to another $z\sim3$ source observed by \citet{shapley16} which has a direct LyC detection and a relatively high inferred LyC escape fraction $>0.5$.

\section{Conclusion} \label{sec:conclusion}
In this work we report the detection of a multiply imaged double-peaked LAE in the MUSE coverage of the RELICS cluster RXC~J0018.5+1626, dubbed RXC0018-LAE1. Unlike most observed double-peaked LAEs, this object shows a peculiar Ly$\alpha$ emission profile dominated by its blue peak rather than the red peak. To the best of our knowledge this is the first observed object showing this kind of Ly$\alpha$ profile at moderate redshifts to date. Using the rest-frame optical [\ion{O}{iii}] and H$\beta$ emission lines we find the systemic redshift of this object to be $z_{\mathrm{sys}}=3.2177\pm0.0001$. Throughout our analysis of RXC0018-LAE1 we find this galaxy to have numerous properties that set it apart from typical LAEs and might indicate it to also leak LyC photons. These properties are:

\begin{itemize}
    \item A unique double-peaked Ly$\alpha$ emission profile in which the blue peak is stronger than the red peak by a blue-to-red ratio of $1.7\pm0.1$.
    \item Possible escape of Ly$\alpha$ photons at the line center.
    \item A relatively large Ly$\alpha$ EW of $\mathrm{EW}_0=(63\pm2)$\,\AA\ and luminosity of $L_{\mathrm{Ly}\alpha}=(6.7\pm0.5)\times10^{42}\,\frac{\mathrm{erg}}{\mathrm{s}}$.
    \item Extremely blue UV continuum slopes and in particular a flattening towards the FUV with $\beta_{\mathrm{NUV}}=-3.0\pm0.2$ and $\beta_{\mathrm{FUV}}=-2.23\pm0.06$.
    \item Relatively low optical nebular line EWs, $\mathrm{EW}_0=(159\pm51)$\,\AA\ for [\ion{O}{iii}]$\lambda5007$\AA\ and $\mathrm{EW}_0=(33\pm12)$\,\AA\ for H$\beta$, an [\ion{O}{iii}]$\lambda5007$\AA/H$\beta$ ratio of $4.8\pm0.7$ and a non-detection of the [\ion{O}{ii}] doublet resulting in a $3\sigma$ lower limit on the [\ion{O}{iii}]/[\ion{O}{ii}] ratio of $>9.3$.
    \item An SED-fit with \texttt{BEAGLE} that shows a low stellar mass $\log(M_{\star}/\mathrm{M}_{\odot})=7.50_{-0.21}^{+0.34}$, a very low dust attenuation $\hat{\tau}_V=0.009_{-0.006}^{+0.010}$ and predicts LyC escape fractions of $f_{\mathrm{esc},\mathrm{LyC}}>0.7$.
\end{itemize}

Given these properties, RXC0018-LAE1 is an interesting target to observe in wavelengths bluer than the Lyman-limit 912\,\AA\ ($<3765$\,\AA\ at $z=3.218$) in order to directly detect its LyC emission and constrain the escape fraction with, e.g., deep HST UV imaging. Further deep high-resolution spectroscopy will also be needed both to resolve the main component's Ly$\alpha$ emission at the line center, to further constrain its rest-frame UV emission lines and to study the two smaller objects in order to determine (i) if they are indeed companions to RXC0018-LAE1 at the same redshift and (ii) how they dynamically interact and how this affects the LyC escape. It is possible that the interaction with the companion galaxies facilitates low \ion{H}{i} column density channels crucial to LyC escape. It would therefore also be beneficial to study the circumgalactic medium of RXC0018-LAE1 with different diagnostics than the resonant Ly$\alpha$ line such as e.g. rest-frame far-infrared (FIR) lines redshifted to \textit{Atacama Large Millimeter/submillimeter Array} (ALMA) bandpass at $z\simeq3.218$. The high magnification factors ($\mu\sim7-10$) and image multiplicity, which allows us to stack observations of both images to gain additional depth, make RXC0018-LAE1 a prime target for deep follow-up observations to study a strong candidate LyC leaking, low-mass high-redshift galaxy analog.

\section*{Acknowledgements}
The authors warmly thank the anonymous referee for their comments and feedback which greatly helped to improve the paper. We further thank Charlotte Mason, Max Gronke and Anja von der Linden for insightful discussions and feedback. LF and AZ acknowledge support by Grant No. 2020750 from the United States-Israel Binational Science Foundation (BSF) and Grant No. 2109066 from the United States National Science Foundation (NSF). AZ acknowledges support by the Ministry of Science \& Technology, Israel. DPS acknowledges support from the National Science Foundation through the grant AST-2109066. MB acknowledges support by the Slovenian National Research Agency (ARRS) through grant N1-0238. This work makes use of observations obtained at the international Gemini Observatory, a program of NSF’s NOIRLab, which is managed by the Association of Universities for Research in Astronomy (AURA) under a cooperative agreement with the NSF. The Gemini Observatory partnership comprises: the NSF (United States), National Research Council (Canada), Agencia Nacional de Investigaci\'{o}n y Desarrollo (Chile), Ministerio de Ciencia, Tecnolog\'{i}a e Innovaci\'{o}n (Argentina), Minist\'{e}rio da Ci\^{e}ncia, Tecnologia, Inova\c{c}\~{o}es e Comunica\c{c}\~{o}es (Brazil), and Korea Astronomy and Space Science Institute (Republic of Korea). Note that the Gemini North Telescope is located within the Maunakea Science Reserve and adjacent to the summit of Maunakea. We are grateful for the privilege of observing the Universe from a place that is unique in both its astronomical quality and its cultural significance. This work is based on observations obtained with the NASA/ESA \textit{Hubble Space Telescope} (HST), retrieved from the \texttt{Mikulski Archive for Space Telescopes} (\texttt{MAST}) at the \textit{Space Telescope Science Institute} (STScI). STScI is operated by the Association of Universities for Research in Astronomy, Inc. under NASA contract NAS 5-26555. Finally, this work is also based on observations made with ESO Telescopes at the La Silla Paranal Observatory obtained from the ESO Science Archive Facility. This research made use of \texttt{Astropy},\footnote{\url{http://www.astropy.org}} a community-developed core Python package for Astronomy \citep{astropy13,astropy18} as well as the packages \texttt{NumPy} \citep{vanderwalt11}, \texttt{SciPy} \citep{virtanen20} and some of the astronomy \texttt{MATLAB} packages \citep{maat14}. The \texttt{Matplotlib} package \citep{hunter07} was used to create the figures in this work. Special thanks also to Eduardo Vitral for lending us some of the coordinate routines of his \texttt{BALRoGO} code \citep[][]{vitral21}.

\section*{Data Availability}
The RELICS HST data and catalogs on which this work is based are publicly available on the \texttt{MAST} archive in the RELICS repository for RXC0018\footnote{\url{https://archive.stsci.edu/missions/hlsp/relics/rxc0018+16/}}. The VLT HAWK-I and MUSE data are publicly available on the ESO Science Archive
\footnote{\url{http://archive.eso.org/scienceportal/home}} under program IDs 0103.A-0871(B) and 0103.A-0777(B) respectively. Finally, the Gemini/GNIRS data and lensing models used in this work will be shared by the authors on request. 


\bibliographystyle{mnras}
\bibliography{references} 


\appendix

\section{Search for further emission lines} \label{app:emission_lines}
Our GNIRS and MUSE spectra yielded clear detections of the [\ion{O}{iii}], H$\beta$ and Ly$\alpha$ lines in RXC0018-LAE1 respectively, as described in sections~\ref{sec:zsys} and~\ref{sec:Lya}. In addition to those, we search the spectra for other emission lines at the wavelengths where they can be expected given the systemic redshift $z_{\mathrm{sys}}=3.2177\pm0.0001$. We derive upper limits and some tentative detections as described in the following sections. Because of their weak signal-to-noise however, we defer a detailed analysis of these possible emission features to future work with deeper spectroscopic observations.

\subsection{The [\ion{O}{ii}] doublet upper limit in GNIRS} \label{app:OII_upper_limit}

\begin{figure}
    \centering
    \includegraphics[width=\columnwidth, keepaspectratio=true]{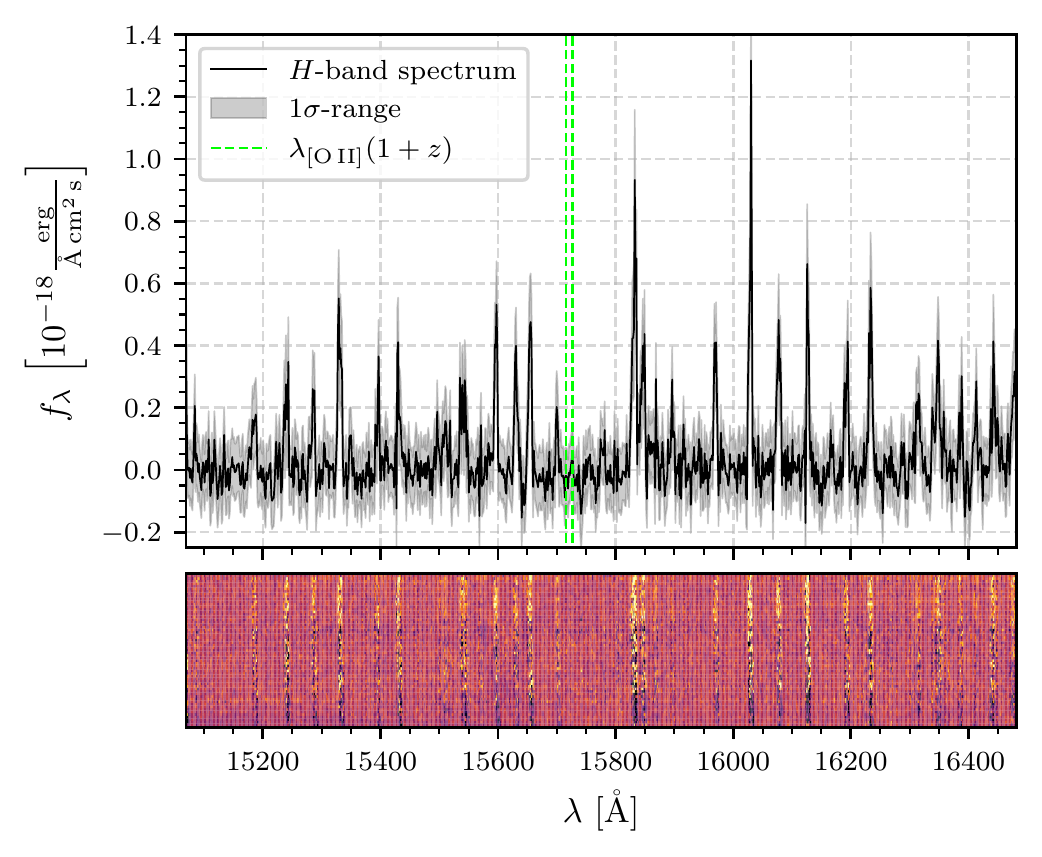}
    \caption{Observed GNIRS \textit{H}-band spectrum of RXC0018-7203 (black) and its $1\sigma$-range (grey shaded area). The green dashed lines show where the [\ion{O}{ii}]$\lambda\lambda3726$\AA,$3729$\AA~doublet can be expected given the spectroscopic redshift of RXC0018-LAE1 measured with the \textit{K}-band lines.}
    \label{fig:H-band_spectrum}
\end{figure}

The \textit{H}-band spectrum of our GNIRS observations of RXC0018-7203 is shown in Fig.~\ref{fig:H-band_spectrum}. While it contains many sky lines, the spectral region where we may expect the [\ion{O}{ii}]$\lambda\lambda3726$\AA,$3729$\AA~doublet at $z\simeq3.218$, $\lambda\simeq15716$\,\AA\ and $\lambda\simeq15729$\,\AA, is relatively free. Nonetheless, we do not detect any significant flux at the expected location of the [\ion{O}{ii}] doublet. Integrating the variance spectrum in a 30\,\AA\ window around the expected wavelengths, we measure the $1\sigma$ upper limit on the total magnification-corrected flux and the rest-frame EW using the \textit{H}-band photometry given in Tab.~\ref{tab:photometry}. The upper limits are shown in Tab.~\ref{tab:UV-lines}.

\subsection{Rest-frame UV lines in MUSE} \label{app:UV_lines}

\begin{table*}
    \centering
    \caption{Rest-frame UV spectroscopy results for RXC0018-LAE1 measured in the MUSE and GNIRS data. The integrated line fluxes were corrected for the gravitational magnification given in Tab.~\ref{tab:photometry}. In the case of non-detections, we quote $1\sigma$ upper limits here. The $\frac{F_{\mathrm{line}}}{\Delta F_{\mathrm{line}}}$ columns contain integrated signal-to-noise ratios.}
    \begin{tabular}{lccccccc}
    \hline                                          &                                           &   \multicolumn{3}{c}{RXC0018-LAE1.1}                                                                      &   \multicolumn{3}{c}{RXC0018-LAE1.2}\\\hline
    Line                                            &   $\lambda_{\mathrm{obs}}$~[\AA] &   $F_{\mathrm{line}}$~$\left[10^{-18}\,\frac{\mathrm{erg}}{\mathrm{s\,cm}^2}\right]$    &   $\mathrm{EW}_0$~[\AA]  &   $\frac{F_{\mathrm{line}}}{\Delta F_{\mathrm{line}}}$   &   $F_{\mathrm{line}}$~$\left[10^{-18}\,\frac{\mathrm{erg}}{\mathrm{s\,cm}^2}\right]$    &   $\mathrm{EW}_0$~[\AA]  &   $\frac{F_{\mathrm{line}}}{{\Delta F_{\mathrm{line}}}}$\\\hline
    \ion{Si}{iv}\,$\lambda1394$\AA       &   -                                       &   $<0.26$        &   $<0.9$         &   -       &   $<0.15$        &   $<0.8$         &   -\\
    \ion{Si}{iv}$\,\lambda1403$\AA       &   -                                       &   $<0.26$        &   $<0.9$         &   -       &   $<0.14$        &   $<0.7$         &   -\\
    \ion{C}{iv}$\,\lambda1548$\AA        &   $6525.6\pm0.5$                 &   $0.40\pm0.10$  &   $1.7\pm0.4$    &   4.0     &   $0.45\pm0.08$  &   $2.8\pm0.5$    &   5.6\\
    \ion{C}{iv}$\,\lambda1551$\AA        &   $6531.2\pm1.7$                 &   $<0.10$        &   $<0.4$         &   -       &   $0.19\pm0.07$  &   $1.2\pm0.5$    &   2.7\\
    \ion{He}{ii}\,$\lambda1640$\AA       &   $6913.4\pm0.4$                 &   $0.37\pm0.10$  &   $1.7\pm0.5$    &   3.7     &   $0.47\pm0.07$  &   $3.6\pm0.4$    &   6.7\\
    \ion{O}{iii}$]\lambda1661$\AA       &   -                                       &   $<0.14$        &   $<0.7$         &   -       &   $<0.09$        &   $<0.7$         &   -\\
    \ion{O}{iii}$]\lambda1666$\AA       &   -                                       &   $<0.14$        &   $<0.7$         &   -       &   $<0.08$        &   $<0.6$         &   -\\
    \ion{C}{iii}$]\lambda1907$\AA       &   $8036.4\pm0.3$                 &   $0.42\pm0.08$  &   $2.7\pm0.5$    &   5.3     &   $0.62\pm0.08$  &   $6.2\pm0.6$    &   7.8\\
    \ion{C}{iii}$]\lambda1909$\AA       &   $8045.1\pm0.4$                 &   $0.39\pm0.07$  &   $2.5\pm0.5$    &   5.6     &   $0.42\pm0.06$  &   $4.2\pm0.6$    &   7.0\\\hline
    $[$\ion{O}{ii}$]\lambda\lambda3726,3729$\AA   &   -                                       &   $<0.07$        &   $<2.3$         &   -       &   -                       &   -                       &   -\\\hline
    \end{tabular}
    \label{tab:UV-lines}
\end{table*}

\begin{figure}
    \centering
    \includegraphics[width=\columnwidth, keepaspectratio=true]{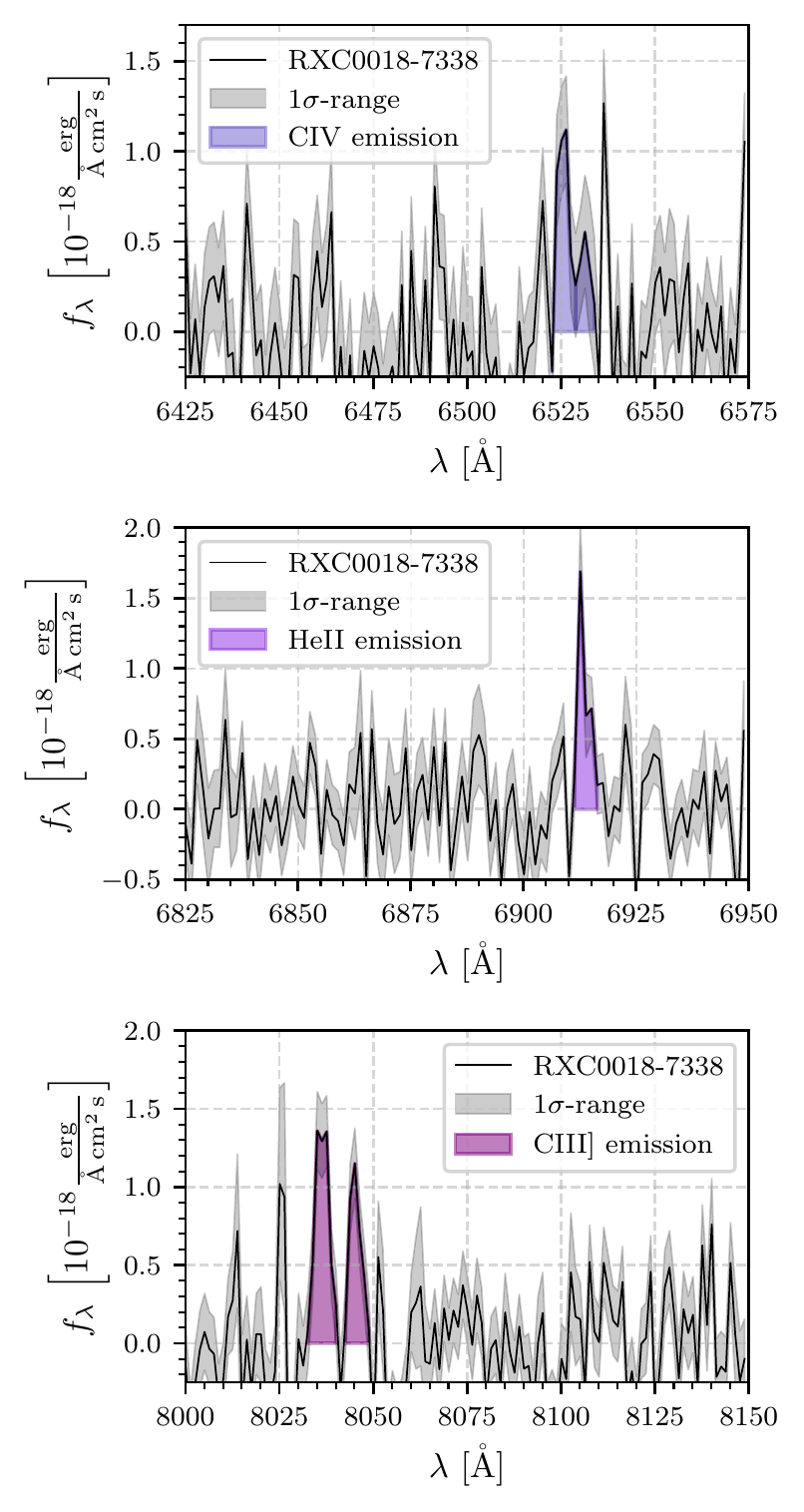}
    \caption{Continuum-subtracted \ion{C}{iv} (\textit{top panel}), \ion{He}{ii} (\textit{middle panel}) and \ion{C}{iii}] (\textit{bottom panel}) emission in RXC0018-7338. The color shaded areas represent the integrated line fluxes and the grey shaded area represents the $1\sigma$-range of the observed spectra.}
    \label{fig:UV-lines}
\end{figure}

In order to search for additional rest-frame UV emission lines beyond the prominent double-peaked Ly$\alpha$ feature in RXC0018-LAE1, we carefully inspect the entire spectral axis of the MUSE cube. A particular criterion for this is that the potential emission features appear consistently in both multiple images. We tentatively find some weak emission in both images at the expected wavelengths of the \ion{C}{iv}\,$\lambda\lambda1548$\AA,$1551$\AA\ and \ion{C}{iii}]$\lambda\lambda1907$\AA,$1909$\AA~doublets and the \ion{He}{ii}$\,\lambda1640$\AA\ line. We do not find the \ion{Si}{iv}$\,\lambda\lambda1394$\AA,$1403$\AA\ and \ion{O}{iii}]$\lambda\lambda1661$\AA,$1666$\AA~doublets where expected given the systemic redshift of our target however. We also do not find any significant emission lines for the tertiary component (IDs 7315 and 7495) consistently over both images.

We extract circular regions of the diameter of the MUSE PSF FWHM (see Fig.~\ref{fig:lya_S/N}) around the main component of RXC0018-LAE1 (i.e. RXC0018-7203 and RXC0018-7338) from the continuum-subtracted MUSE cube constructed in section~\ref{sec:flux_extraction} and collapse them to 1D spectra. We then use the same methods as described in sections~\ref{sec:zsys} and~\ref{sec:Lya_properties} to measure line centers, integrated flux and rest-frame EWs of these lines. The measured properties can be found in Tab.~\ref{tab:UV-lines} and we show the \ion{C}{iv}, \ion{He}{ii} and \ion{C}{iii}] spectra of RXC0018-7338 in Fig.~\ref{fig:UV-lines}. We measure upper limits for the lines that we do not detect by integrating the variance spectrum in a 20\,\AA\ window around the expected line position. The EWs in Tab.~\ref{tab:UV-lines} are computed using the UV-continuum fits obtained in section~\ref{sec:UV_continuum} to estimate the continuum.

As is apparent from the values in Tab.~\ref{tab:UV-lines}, these detections are very weak. Indeed, the peak flux density barely exceeds $2\sigma$, in particular for the \ion{C}{iv} lines (i.e. the integrated flux might be higher as can be seen in Tab.~\ref{tab:UV-lines}, but the flux density in each spectral element does not exceed $\sim2\sigma$). A recent study of low-redshift LyC leakers has established a tentative connection between the LyC escape fraction and the \ion{C}{iv}/\ion{C}{iii}]-doublet flux ratio \citep{schaerer22}. Summing the fluxes of each doublet, we find a \ion{C}{iv}/\ion{C}{iii}]-ratio of $0.6\pm0.1$ based on the measurement for RXC0018-LAE1.2 (see Tab.~\ref{tab:UV-lines}) which is just below the ratios typically found for strong LyC leakers in \citet{schaerer22}, i.e. $\gtrsim0.7$. Note that this estimate is based on only 3 galaxies though and given the uncertainties in both our and the \citet{schaerer22} measurements, a high LyC escape fraction would still be consistent in RXC0018-LAE1. Moreover, as can be seen in Fig.~\ref{fig:UV-lines}, our detections are rather tentative though, in particular that of \ion{C}{iv}\,$\lambda1551$\AA, even in the more magnified image. We will therefore need deeper spectroscopy to obtain more robust detections to better measure this ratio. Note that the same study by \citet{schaerer22} found the presence of \ion{He}{ii}$\,\lambda1640$\AA\ to also correlate with LyC escape since it indicates a strong ionization field.

Interestingly, the \ion{He}{ii}$\,\lambda1640$\AA\ seems to be spatially shifted towards the secondary component (IDs 7202 and 7338) in both multiple images. While the effective spatial resolution of the MUSE data (see e.g. Fig.~\ref{fig:lya_S/N} for the FWHM of the MUSE PSF) does not allow us to properly resolve the exact spatial origin of the emission line, this might nonetheless indicate that the \ion{He}{ii} emission actually originates from the secondary component rather than the main component like the other emission lines measured in this work. If this were indeed the case, it would place the secondary component at a spectroscopic redshift $z=3.2156\pm0.0003$ which would make it a close companion to the Ly$\alpha$-emitting galaxy. Note that the difference between this measurement and the systemic redshift of the main component is not sufficient to account for the peak separation of the Ly$\alpha$ emission. Our previous conclusions that the double-peaked Ly$\alpha$ originates from one and the same object, the main component of RXC0018-LAE1, are therefore not contradicted by this detection. We will need deeper observations with higher spectral and spatial resolution to robustly detect these tentative emission lines, measure their properties and determine if the \ion{He}{ii} emission indeed originates from the secondary component.


\bsp	
\label{lastpage}
\end{document}